\documentclass[fleqn,aps,prl,twocolumn]{revtex4}

\usepackage[]{amsmath}
\usepackage{amssymb}
\usepackage[dvips]{graphicx}
\usepackage{color}
\usepackage{tabularx}
\usepackage{mathtools}
\usepackage{algpseudocode}
\usepackage{enumitem}
\usepackage{multirow}
\usepackage{epstopdf}  
\usepackage{bm}
\usepackage{hyperref}
\usepackage[version=3]{mhchem}
\usepackage{tikz}
\usetikzlibrary{shapes,snakes}
\usetikzlibrary{arrows}


\begin{document}

\title{
  The Phase Diagram of a Deep Potential Water Model
  }
\author{Linfeng Zhang}
\affiliation{Program in Applied and Computational Mathematics, 
Princeton University, Princeton, NJ 08544, USA}
\author{Han Wang}
\email{wang\textunderscore han@iapcm.ac.cn}
\affiliation{Laboratory of Computational Physics,
  Institute of Applied Physics and Computational Mathematics, Fenghao East Road 2, Beijing 100094, P.R.~China}
\author{Roberto Car}
\email{rcar@princeton.edu}
\affiliation{Department of Chemistry, Department of Physics, Program in Applied and Computational Mathematics, Princeton Institute for the Science
and Technology of Materials, Princeton University, Princeton, New Jersey 08544, USA}
\author{Weinan E}
\affiliation{Department of Mathematics and Program 
in Applied and Computational Mathematics, 
Princeton University, Princeton, NJ 08544, USA}
\affiliation{Beijing Institute of Big Data Research, 
Beijing, 100871, P.R.~China}

\begin{abstract}
Using the Deep Potential methodology, we construct a model that reproduces accurately the potential energy surface   
of the SCAN approximation of density functional theory for water, from low temperature and pressure to about 2400 K and 50 GPa, excluding the vapor stability region. The computational efficiency of the model makes it possible to predict its phase diagram using molecular dynamics. Satisfactory overall agreement with experimental results is obtained. The fluid phases, molecular and ionic, and all the stable ice polymorphs, ordered and disordered, are predicted correctly, with the exception of ice III and XV that are stable in experiments, but metastable in the model. The evolution of the atomic dynamics upon heating, as ice VII transforms first into ice VII$''$ and then into an ionic fluid, reveals that molecular dissociation and breaking of the ice rules coexist with strong covalent fluctuations, explaining why only partial ionization was inferred in experiments.   

\end{abstract}

\maketitle

The phase diagram of water is extremely rich. 
In the temperature and pressure domain with  $T \lessapprox 400$~K and  $P \lessapprox 50$~GPa, there are ten stable phases, nine solid (ice Ih, II, III, V, VI, VII, VIII, XI and XV) and one liquid, in addition to five metastable phases (ice IV, IX, XII, XIII and XIV)~\cite{bridgman1912water,bridgman1937phase,salzmann2009ice}.
This large variety of structures are made possible by hydrogen bonded arrangements of the molecules. 
In ice, the oxygen sublattice is crystalline, but the hydrogen sublattice can be either ordered or disordered, due to the vast number of nearly degenerate hydrogen (proton) configurations allowed by the ice rules. 
The corresponding configurational or residual entropy stabilizes disordered polymorphs at high temperature. 
Thus, near melting, all the stable phases are disordered (ice Ih, III, V, VI, and VII). 
In ice Ih, VI, and VII, disorder is complete and the residual entropy is well approximated by $k_B\ln 1.5 \approx 0.4055 k_B$/\ce{H_2O}~\cite{pauling1935structure}. 
In ice III and V, disorder is partial~\cite{londono1993neutron,lobban2000structure} and the entropy is less than Pauling's estimate but still significant~\cite{macdowell2004combinatorial}. 
Upon cooling, ice Ih, VI, and VII become less stable than their ordered counterparts, ice XI~\cite{yen2015proton}, XV~\cite{salzmann2009ice}, and VIII~\cite{pruzan1994pressure}, respectively. 
Ordered polymorphs are ferroelectric (XI) or anti-ferroelectric (II, XV, and VIII). 
Interestingly, ice II does not have a disordered counterpart. See, e.g.,~Ref.~\cite{salzmann2011polymorphism} for a review of ice polymorphism.      

At high pressure, the stability of the solid phases extends to higher temperatures, the hydrogen bonds weaken, and molecular dissociation into ions is promoted by the increasing thermal fluctuation. 
Molecular to ionic transformation is continuous in the fluid. 
In the solid, for  $T \gtrapprox 850 $~K and pressures above $\approx$ 14 GPa, ice VII transforms into ice VII$''$, a superionic phase in which the BCC oxygen sublattice of ice VII coexists with mobile protons. 
Upon further heating, ice VII$''$ melts into an ionic fluid~\cite{cavazzoni1999superionic,goldman2005bonding,queyroux2020melting,hernandez2016superionic,hernandez2018proton}.        

Molecular dynamics (MD) simulations give microscopic insight into the water phases and complement experiments with atomistic details~\cite{sanz2004phase,abascal2005general,abascal2005potential,vega2005can,aragones2009plastic,chen2017ab,rozsa2018ab}. 
The key ingredient of MD is the potential energy surface (PES), 
which can be constructed either by fitting a physically motivated force field to experiment, or, non-empirically, from quantum theory ({\it ab initio} MD (AIMD)). 
Comparing the phase diagram predicted by MD to experiment is the ultimate accuracy test of a model PES. 
Due to the high computational cost of AIMD, extensive studies of the water phase diagram have only been possible so far with empirical force fields, which, however, face severe difficulties with the ionic phases. 
By contrast, in AIMD, the PES is constructed on-the-fly from density functional theory (DFT) and can describe molecular dissociation processes. 
Indeed, this approach has been particularly useful in modeling proton transfer in the liquid at ambient conditions~\cite{marx1999nature,chen2017ab}, or the superionic ice phases at high pressure and temperature~\cite{hernandez2016superionic}.

Advances in machine learning (ML) are making possible MD simulations of {\it ab initio} quality at a cost of empirical force fields. 
Applications to water studied the phase behavior at ambient~\cite{morawietz2016van,cheng2019ab} and deeply undercooled~\cite{gartner2020signatures} conditions, isotopic effects~\cite{cheng2016nuclear,cheng2019ab,ko2019isotope}, infrared and Raman spectra~\cite{gastegger2017machine,raimbault2019using,zhang2020deep,sommers2020raman}, etc.    
{A recent calculation reported the phase diagram in the ($T,P$) range from $150$~K to
$300$~K and from $0.01$~GPa to $1$~GPa, at the hybrid DFT level, including nuclear quantum effects~\cite{reinhardt2021quantum}}.
However, to the best of our knowledge, no attempt has been made to describe water in a wide thermodynamic range including ordered and disordered ice, superionic ice, molecular and ionic fluid phases.   

Here this goal is achieved with Deep Potential Molecular Dynamics (DPMD)~\cite{zhang2018deep,zhang2018end}, using
an iterative concurrent learning scheme, Deep Potential (DP) Generator~\cite{zhang2019active,zhang2020dp}, to construct the PES with SCAN-DFT as the reference. SCAN~\cite{sun2015strongly} is a non-empirical functional that describes well several properties of water~\cite{sun2016accurate}.
We find that a unique DP model can reproduce closely DFT in a vast thermodynamic range, extending from ambient pressure to $\approx$ 50 GPa and from $\approx$ 50 K to $\approx$ 2000 K, excluding the vapor stability region. 
DPMD predicts the stable phases, including ordered, disordered, and superionic ices, as well as molecular and ionic fluid phases.
Overall, the phase diagram agrees well with experiment, further validating the quality of the SCAN approximation. In the high $(T,P)$ region the simulations reveal key features of the temperature induced transitions from ice VII to ice VII$''$ and from the latter to an ionic fluid.   

To construct the model PES, a trial DP is built from configurations of the liquid, at ambient conditions, and of all the experimentally known stable and metastable ice polymorphs for $P \lessapprox 50 $~GPa (Ih, Ic, II, III, IV, V, VI, VII, VIII, IX, XI, XII, XIII, XIV, XV). The model is used by DP Generator to explore a wide region of the phase space with isothermal-isobaric ($NPT$) DPMD trajectories. 
The protocol is iterated to refine the model with new DFT data until satisfactory accuracy is achieved. 
The visited states can be roughly classified into three groups: the low pressure (A), the high pressure (B), and the superionic group (C). Group (A) includes states in the range $50 \leq T \leq 600$~K and $10^{-4}\leq P \leq 5$~GPa,
starting from configurations of the fluid and of all the ices 
except VII and VIII. 
Group (B) includes states in the range $50\leq T \leq 600$~K and $0.1\leq P \leq 50$~GPa,
starting from configurations of ice VII and VIII. 
Group (C) includes states in the range $200\leq T \leq 2400$~K and $1\leq P \leq 50$~GPa,
starting from ice VII and the fluid. 
DPMD samples almost uniformly the thermodynamic domains of the three groups.  
The deviation in the predicted forces within a set of representative DP models is used to label
configurations for which new DFT calculations of the energy, forces, and virial are necessary.
The new data are added to the training dataset and serve to refine the representative DP models entering the next iteration.

After 36 concurrent learning iterations the error in the force is satisfactorily reduced and the procedure ends. 
The accumulated number of snapshots in the training dataset is 31058, a tiny fraction ($\sim0.05\%$) 
of the configurations visited by DPMD.
At this point, the relative energies of configurations within each phase are well described, but deviations from DFT still affect the averages. 
To reduce these deviations below a  small threshold, 3519 additional training configurations are necessary~\footnote{Model and data publicly available at DP Library~\url{http://dplibrary.deepmd.net}.} (Supplementary Material (SM) Fig.~S1).

The Vienna {\it ab initio} simulation package (VASP) version 5.4.4~\cite{kresse1996efficiency,kresse1996efficient} is used for the DFT calculations. DeePMD-kit~\cite{wang2018deepmd} is used for DP training and for running DPMD, interfaced with  LAMMPS~\cite{plimpton1995fast}.
DP-GEN~\cite{zhang2020dp} is used for the concurrent learning process. See details
in SM.


{\it Accuracy of the DP model.}
The error relative to DFT is quantified with an independent testing dataset including 5141 configurations along 67 isothermal-isobaric DPMD trajectories spanning the relevant thermodynamic domain (SM Fig.~S2). 
In most cases, the root mean square error (RMSE) of energy and force is $\sim 1$meV/\ce{H_2O} and $\sim50$meV/$\textrm{\AA}$, respectively. 
Larger absolute errors may be possible at high temperature, but since thermal fluctuations are large, the relative RMSE is still of $\sim 10\%$ or less.

{\it Phase diagram.}
Thermodynamic integration is used to compute the absolute Gibbs free energy of a single state point of each phase~\cite{frenkel1984new,vega2008determination}.
The algorithm of Ref.~\cite{buch1998simulations} is used to generate the fully disordered structures of ice Ih, IV, VI, and VII, and Pauling's residual entropy contribution ($0.4055 k_B$/\ce{H_2O}) is added to their free energy.
For the partially disordered structures of ice III and V and the corresponding entropies we follow Ref.~\cite{macdowell2004combinatorial}.
To minimize finite size effects we use cells with at least 128 molecules. 
Taking into account finite size, entropy approximation, DP error, and statistical uncertainty we estimate that the free energy error should be approximately 1~meV/\ce{H_2O}.
Then, using thermodynamic integration with the composite Simpson rule we trace a family of curves representing the variation
with pressure along an isotherm, or with temperature along an isobar, of the free energy of each phase. 
The intersections between pairs of curves define phase coexistence points.  
Finally, the phase boundary lines stemming from the coexistence points are traced by integrating the Gibbs-Duhem equation~\cite{kofke1993gibbs} with a second order Runge-Kutta method. See SM,~Sec.~SIII~A. 


The numerical accuracy of the predicted phase boundaries can be gauged from the 
consistency of the predicted triple points (TPs). Each TP can be inferred in three independent ways
from the intersection of two boundary lines between the three coexisting phases. 
The average of these estimates defines a TP, and the 
standard deviation gives the estimated error. 
From the TPs in 
SM~Tab.~SV,
we infer that the numerical uncertainty of the calculated phase boundaries is less than 5~K in temperature and less than 0.02~GPa
in pressure. 

\begin{figure}
  \centering
   \includegraphics[width=0.48\textwidth]{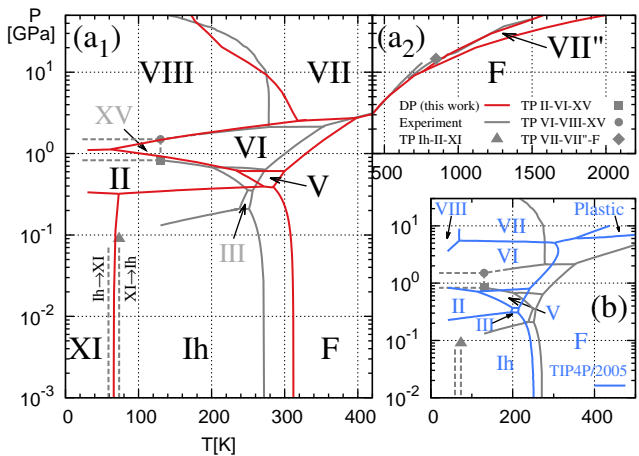}
  \caption{Phase diagram of water.
  (a$_1$) DP model (red solid lines) and experiment (gray solid lines) for $T<420$ K. 
  Black letters indicate phases that are stable in experiment and model. 
  Ice III and XV (stable in experiment but metastable in model) are gray. 
  Experimental coexistence lines 
  are from Ref.~\cite{wagner2011new} (melting curves), and from Refs.~\cite{brown1966preliminary,pruzan1994pressure,journaux2020large} (solid-solid curves).
  The gray triangle indicates the postulated Ih-II-XI TP~\cite{journaux2020large}.
  The two dashed lines indicate the experimentally observed transitions Ih$\rightarrow$XI and XI$\rightarrow$Ih~\cite{yen2015proton}.
  The gray solid circle and square denote the VI-VIII-XV and the II-VI-XV TPs, respectively~\cite{salzmann2009ice}.
  (a$_2$) Phase diagram at high $T$ and $P$.  
  The experimental melting lines are from Ref.~\cite{datchi2000extended} 
  and Ref.~\cite{queyroux2020melting}. The
  VII-VII$''$-F TP is from Ref.~\cite{queyroux2020melting}.
  (b) Phase diagram of TIP4P/2005 water~\cite{aragones2009phase}.
  }
  \label{fig:phase}
\end{figure}

Overall, the DP phase diagram in Fig.~\ref{fig:phase}~(a) agrees well with experiment.
All the stable ice phases 
are predicted correctly, with two exceptions, ice III and XV, 
which are metastable in the DP model. 
The Ih-F coexistence line is displaced by $\approx$ 40 K to higher temperature than experiment, while the Ih-II line is displaced by $\approx$ 0.02 GPa to higher pressure than experiment. 
Thus, the stability of Ih is overestimated, consistent with the tendency of the SCAN approximation to overestimate the hydrogen bond strength~\cite{sun2016accurate}. 
On the other hand, the Ih-XI boundary is predicted correctly, reflecting the close similarity of the hydrogen bond configurations in the two systems. 
The shift to higher pressure of the Ih-II boundary may contribute to the metastability of ice III. 
The metastability of ice XV may reflect a general difficulty of gradient corrected functionals to predict the ground state structure of this ice form~\cite{salzmann2009ice}. 
Within the accuracy of the DP model, competing phases differing in free energy by $\lessapprox 1$~meV/\ce{H_2O} should be considered degenerate. 
This happens to IV and VI in part of the stability domain of the latter (SM~Sec.SIII~B). 
The coexistence lines for $P \gtrapprox 1$~GPa including the ice-fluid boundary, the  VII-VII$''$ boundary, and the VII-VII$''$-F TP 
are also in good qualitative agreement with experiment. 
At pressures higher than reported in Fig.~\ref{fig:phase}, ice VII transforms into ice X~\cite{polian1984new}. 
This regime is beyond the domain of validity of the present DP model and is not investigated.

It is instructive to compare the DP phase diagram with the one derived from one of the most accurate empirical water models,
TIP4P/2005 ~\cite{abascal2005general}, which 
assumes rigid molecules and is parameterized with experimental observations, such as, e.g., the temperature of maximal liquid density at ambient pressure, the densities of ice II, III, and V at different thermodynamic conditions, etc.
As shown in Fig.~\ref{fig:phase}~(b), TIP4P/2005 works well at low and intermediate pressures.
At higher pressures, however, significant deviations from experiment affect the boundary lines between ice VIII, VII, and VI. 
Moreover, the rigid molecule approximation does not allow ionized water configurations. At high pressure and temperature
TIP4P/2005 predicts a first-order transition from ice VII to a plastic phase, in which the BCC oxygen sublattice coexists with freely rotating molecules~\cite{takii2008plastic,aragones2009phase}. No experimental evidence has been found so far for this phase, nor was such behavior observed in our DP simulations.

\begin{figure}
    \centering
    \includegraphics{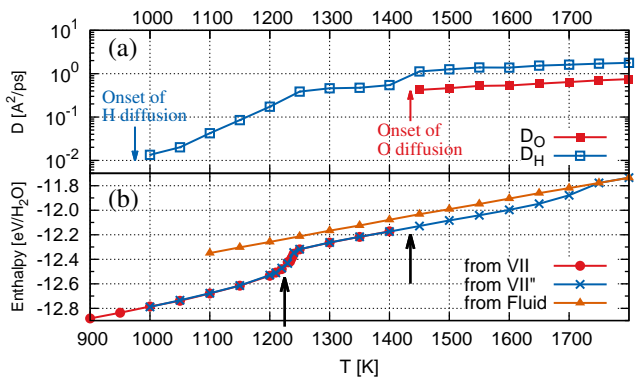}
    \caption{Two-step melting of ice VII along the $P=30$~GPa isobar, showing a solid-solid (VII-VII$''$) followed by a solid 
    fluid (VII$''$-F) transition. 
    (a) H (blue open square) and O (red solid square) diffusion coefficients as a function of temperature.
    (b) Enthalpy as a function of temperature. The estimated temperatures of solid-solid and solid-fluid transitions are indicated by the black arrows.
    }
    \label{fig:ice07-trans}
\end{figure}

\textit{ Ionic phases.}
According to the DP model, at low $T$, ice VII is a molecular crystal with full proton disorder and insignificant atomic diffusion. Upon heating, however, H diffusion grows exponentially with $T$, while O diffusion remains insignificant. 
This behavior is illustrated in Fig.~\ref{fig:ice07-trans}(a) for the isobar at 30 GPa. 
Eventually H diffusion saturates and remains approximately constant over a finite $T$ interval. 
At even higher $T$ the diffusivities of H and O jump to distinct macroscopic values signaling transformation to a fluid.
Ice VII has been referred to as ice VII$'$ and as ice VII$''$ in the thermodynamic domains of exponential growth and saturation of the H diffusivity~\cite{hernandez2016superionic}. 
The enthalpy evolution along the 30 GPa isobar is depicted in Fig.~\ref{fig:ice07-trans}(b). 
It shows a smooth reversible variation in ice VII$'$ followed by a more rapid change when ice VII$'$ turns into ice VII$''$. 
This affects enthalpy and volume, and occurs spontaneously without apparent hysteresis over timescales of 100 ps in simulations with 438 molecules. Simulations with up to 3456 molecules show a sharpening of the rapid change (SM Fig.~S5), as expected of a weakly first-order phase transition. 
Since we cannot associate a separate thermodynamic phase to ice VII$'$, we retain for it the name of ice VII hereafter and in the phase diagram. 
The VII-VII$''$ transition shows the typical behavior of a type II superionic transition~\cite{boyce1979superionic,schwegler2008melting}.
A jump in O diffusivity signals melting of ice VII$''$, a transition 
undetectable by monitoring the enthalpy on timescales of 100 ps in a heat-until-melt simulation. 
Thus, to determine the melting temperature we used a two-phase simulation of 1728 molecules at $50$ GPa, and extrapolated
the melting temperature to lower $T$ with Gibbs-Duhem integration. The predicted VII-VII$''$-F TP is located at (774~K 10.6~GPa), in relatively good agreement with the most recent experimental result (850~K 14.6~GPa)~\cite{queyroux2020melting}. 
Importantly, the same experiment confirmed the first-order nature of the VII-VII$''$ transition, signaled by a discontinuous change of the lattice parameter in X-ray diffraction (XRD). 

\begin{figure}
    \centering
    \includegraphics[width=0.45\textwidth]{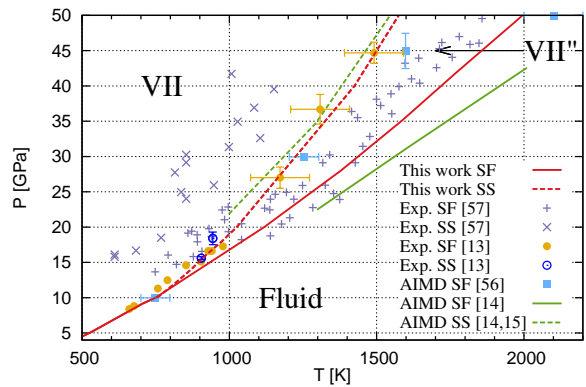}
    \caption{Phase diagram in the superionic region. SF and SS indicate VII$''$-F solid-fluid and VII-VII$''$ solid-solid phase transitions, respectively. Solid and dashed lines indicate the SF and SS coexistence lines according to this work (red) and an earlier AIMD simulation (green). The solid green line is an upper bound for the melting $T$.}
    \label{fig:phase-si}
\end{figure}

A magnified view of the DP phase diagram in the VII-VII$''$-F domain is shown in Fig.~\ref{fig:phase-si}, together with experimental and AIMD results.
Overall there is good qualitative agreement:
experiments confirm the presence of two first-order transitions, a solid-solid and a solid-fluid one. 
Possibly, the significant scatter in the experimental data reflects the difficulty of detecting weakly first-order phase transitions at challenging thermodynamic conditions. 
A two-step melting process for ice VII, with a superionic intermediate, was first proposed in Ref.~\cite{cavazzoni1999superionic} based on AIMD simulations. 
The corresponding solid-solid phase transition was confirmed experimentally in Ref.~\cite{schwager2008h2o}, without structural details on the new solid phase. 
These were provided recently by XRD experiments that verified the BCC lattice structure of VII$''$~\cite{queyroux2020melting}. 
The DP results are in semi-quantitative agreement with AIMD simulations for the VII-VII$''$ and VII$''$-F boundaries~\cite{hernandez2016superionic,hernandez2018proton,schwegler2008melting}.
The differences between DPMD and these earlier studies should be attributed mainly to the adopted exchange-correlation functionals 
and to the relatively small size and time scales of the AIMD simulations.

DP simulations give insight on the atomistic processes that underlie the two-step transition from ice VII to ionic fluid. 
The 
O-O, O-H, and H-H pair correlation functions along the $30$ GPa isobar shown in SM Fig.~S4 
illustrate the progressive loss of long-range order 
as the system progresses through ice VII, VII$''$ and ionic fluid.
Interestingly, in spite of the large diffusivity of H in ice VII$''$ and of H and O in the fluid, the running O-H coordination number retains a well defined shoulder at a value equal to 2, indicating that strong covalent fluctuations favoring neutral water molecules remain effective in presence of ionization and breaking of the ice rules. 
The O sublattice is BCC in ice VII and VII$''$. 
Thus, in ice VII, before the onset of H diffusion each O has 8 O nearest neighbors along the half-diagonals of a cube, 4 of which are occupied by an H atom satisfying the ice rules and 4 of which are empty. 
Upon heating, ice rule breaking fluctuations occur, in which the H atoms oscillate along a bond creating OH$^-$-OH$_3^+$ defect pairs that either rapidly recombine or dissociate as the defects move further apart via Grotthuss-like mechanisms~\cite{marx2010aqueous}. 
A rapid increase of the proton mobility with $T$ follows defect pairs dissociation. 
This process is accompanied by partial occupation of the empty O-O bonds due to molecular rotations, which occur along specific directions and are far from the free rotations hypothesized for the plastic phase. 
As a consequence, the H population of the empty bond network increases, that of the occupied bond network decreases, and the overall H diffusion increases. 
The occupation of interstitial sites outside the bonds remains negligible throughout. 
This trend continues until all the O-O bonds are equally occupied and ice VII transforms to ice VII$''$,
a process marked by a saturation of the H diffusivity and a concomitant volume expansion due to diminished hydrogen bonding forces. 
Proton diffusion is associated to rapid hops along the bonds with Grotthuss like mechanisms not only in ice VII$''$ but also in the ionic fluid.
{The average population of ionic defects at 30 GPa is approximately 7.0 percent at 1250~K in ice VII$''$, and becomes 10.8 percent at 1450~K in the fluid.}
Thus, full ionization is never achieved at these pressures, in agreement with experiment~\cite{nellis1988nature}.

In conclusion, we have shown that DP has made it possible to predict the phase diagram of water from {\it ab initio} quantum theory, over a vast range of temperatures and pressures. 
With further training the potential constructed here could be extended to other thermodynamic conditions, including the vapor and phases at higher temperatures and pressures. 
Extensions to model solutions and interfacial water~\cite{natarajan2016neural,wohlfahrt2020ab,andrade2020free} are also possible.
Competing stable and metastable phases may have free energies within 1 meV/\ce{H_2O} or less, posing a severe challenge both to the accuracy required from the reference quantum model, and to the faithfulness of its neural network representation.
Here we adopted the SCAN approximation of DFT in view of its good balance of efficiency and accuracy, but more accurate functional approximations and/or higher level quantum chemical methods would be possible, in principle. 
Finally, the present study was entirely based on classical MD simulations, but it is known that nuclear quantum effects are responsible for the observed isotopic shifts in the thermodynamic properties of water. 
These shifts are typically smaller than the deviations from experiment of the present classical formulation. 
In future studies one can include these effects
using path integral MD methods, as done, e.g., in Ref~\cite{cheng2019ab,ko2019isotope,reinhardt2021quantum}.


\textbf{Acknowledgement}

The work of H.W. is supported by the National Science Foundation of China under Grant No.11871110 and Beijing Academy of Artificial Intelligence(BAAI).
We thank the Center Chemistry in Solution and at Interfaces (CSI) funded by the DOE Award DE-SC0019394 (L.Z., R.C. and W.E), as well as a gift from iFlytek to Princeton University and the ONR grant N00014-13-1-0338 (L.Z. and W.E).


\end{document}